\documentclass[a4paper, 11pt]{article}
\usepackage[a4paper, total={7in, 10in}]{geometry}
\usepackage[utf8]{inputenc}
\usepackage[cmex10]{amsmath}
\usepackage[pdftex]{graphicx}
\usepackage{hyperref}
\usepackage{parskip}
\usepackage[backend=biber, style=authoryear, natbib=true, maxbibnames=4, maxcitenames=2]{biblatex}
\DeclareNameAlias{sortname}{family-given}

\DefineBibliographyStrings{english}{in = {{}{}}}

\hypersetup{pdfinfo={
	Title={On Determining the Distribution of a Goodness-of-Fit Test Statistic},
	Subject={On Determining the Distribution of a Goodness-of-Fit Test Statistic},
	Author={Sean van der Merwe},
	Keywords={Bayes, Distribution, Gamma, GPD, Hypothesis Testing, Objective Bayes, p-value, Predictive Posterior, Simulation} } }

\newlength{\figwidth} 

\usepackage{authblk}
\title{On Determining the Distribution of a Goodness-of-Fit Test Statistic}
\author[a,*]{Sean~van~der~Merwe}
\affil[a]{University of the Free State, Box 339, Bloemfontein, 9300, South Africa; vandermerwes@ufs.ac.za; +274013770}
\affil[*]{Corresponding author}

\date{2014/05/12}

\begin{document}

\maketitle

\begin{abstract}
We consider the problem of goodness-of-fit testing for a model that has at least one unknown parameter that cannot be eliminated by transformation. Examples of such problems can be as simple as testing whether a sample consists of independent Gamma observations, or whether a sample consists of independent Generalised Pareto observations given a threshold. Over time the approach to determining the distribution of a test statistic for such a problem has moved towards on-the-fly calculation post observing a sample. Modern approaches include the parametric bootstrap and posterior predictive checks. We argue that these approaches are merely approximations to integrating over the posterior predictive distribution that flows naturally from a given model. Further, we attempt to demonstrate that shortcomings which may be present in the parametric bootstrap, especially in small samples, can be reduced through the use of objective Bayes techniques, in order to more reliably produce a test with the correct size.
\end{abstract}


\section{Introduction} \label{sec:intro}

\subsection{Distribution tests where the null model is completely specified} \label{sec:nullspecified}
Well-known tests for determining whether a sample could have arisen from a specific distribution include the Kolmogorov-Smirnov (KS), Anderson-Darling (AD) and similar tests based on the empirical distribution function (see Darling, 1957 for a historical introduction); however, these tests in their base form assume that all parameters in the null model are known.

Specifically, let $\mathbf{X}$ be an i.i.d. random sample of size $n$ from an unknown distribution, and let $S=S(\mathbf{X}|m)$ be a test statistic for testing the null hypothesis that $\mathbf{X}$ follows a specific distribution. In general, the test statistic $S$ depends on the parameters $\boldsymbol\theta$ of the distribution to be tested, as is the case for the KS and AD statistics. Thus, $S=S(\mathbf{X}|\boldsymbol\theta,m)$ is a function of $\boldsymbol\theta$ in general, so that in their base form many distribution tests are suitable for testing a null hypothesis of the form: $\mathbf{X}$ follows a specific distribution with parameters $\boldsymbol\theta_0$ fixed. The test statistic used for testing a null-hypothesis of this form is then $S(\mathbf{X}|\boldsymbol\theta_0,m)$.

Note that for fixed (or known) $\boldsymbol\theta_0$, the exact distribution of $S(\mathbf{X}|\boldsymbol\theta_0,m)$ can often be determined, if necessary by simulation. Hence, we can have an exact test of the null hypothesis that $\mathbf{X}$ follows a specific distribution with parameters $\boldsymbol\theta_0$. By the test being exact, in the classical sense, we mean that the test has correct type 1 error, that is, for a given significance level $\alpha^*$ (say) the test falsely rejects the null hypothesis with probability $\alpha^*$. This condition is equivalent to the p-value having a Uniform distribution under the null hypothesis.

\subsection{Parameters of null model not specified} \label{sec:nullunspecified}

In practice it is often of interest to test whether $\mathbf{X}$ follows a specific distribution, but without specifying the parameters $\boldsymbol\theta$ of the distribution. If, nevertheless, the relevant test statistic $S=S(\mathbf{X}|\boldsymbol\theta,m)$ is a function of $\boldsymbol\theta$, then $S$ might be calculated as $S=S(\mathbf{X}|\hat{\boldsymbol\theta},m)$, where $\hat{\boldsymbol\theta}$ is some estimate of $\boldsymbol\theta$. When doing so in general, the problem arises that the distribution of the test statistic $S(\mathbf{X}|\hat{\boldsymbol\theta},m)$ might not only depend on the distribution family and sample size, but also on the values of the unknown parameters $\boldsymbol\theta$ (\cite[102]{DAgostino1986}, \cite{Darling1957}). The distribution of the test statistic might even be affected by the method of estimation of the unknown parameters.

\begin{figure}
\centering
\includegraphics[width=0.8\figwidth]{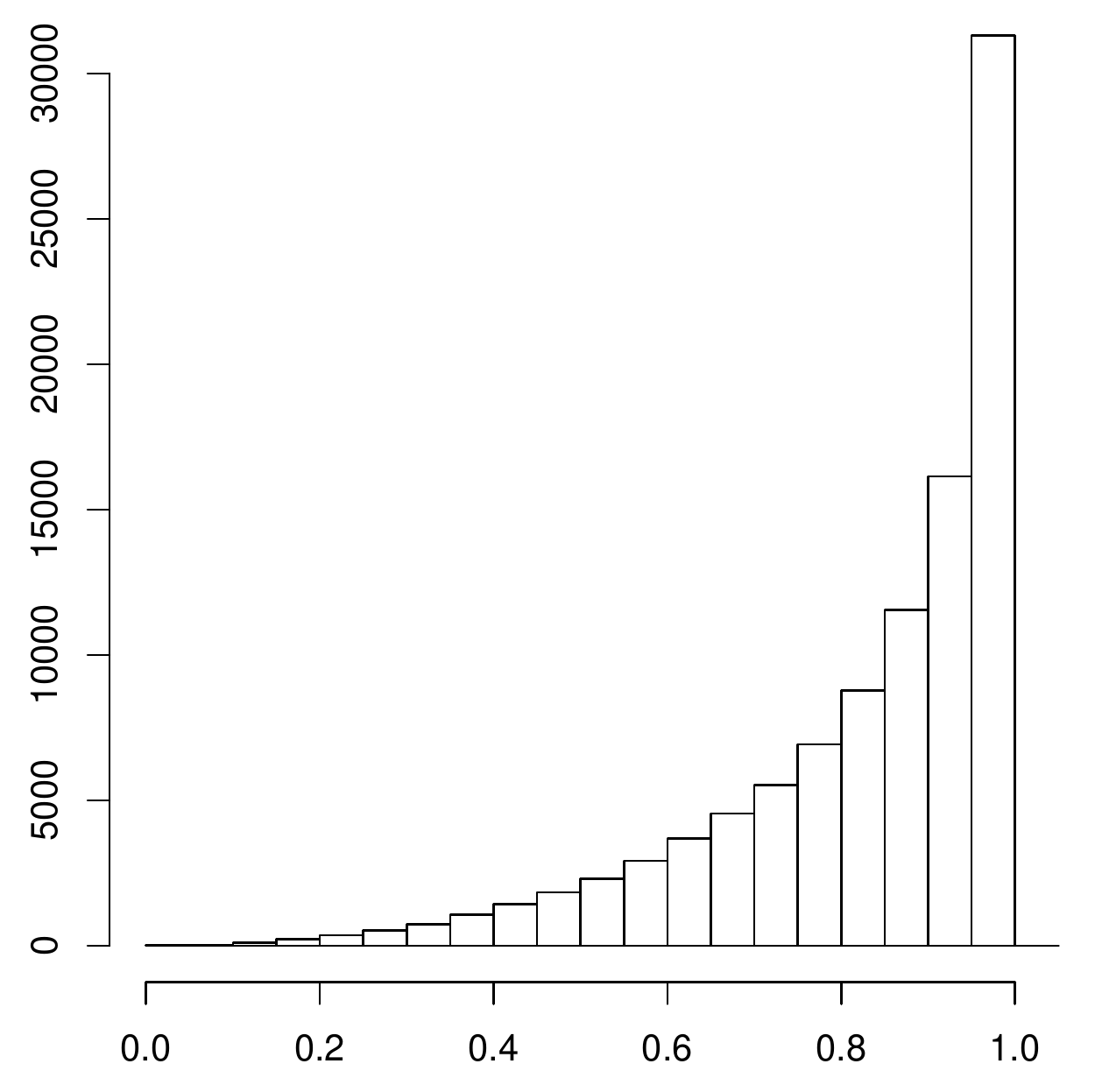}
\caption{Calculated p-values from Gamma samples using Gamma test of \textcite{Marsaglia2004}.} \label{fig:mar}
\end{figure}

To illustrate this problem 100,000 samples from a $Gamma(6,2)$ distribution were simulated and the Gamma test of \textcite{Marsaglia2004} in the ADGofTest package \parencite{ADGofTest} in the statistical software R \parencite{RCore} was performed. This test is based on the principle of replacing the unknown parameters of the Gamma distribution by their estimates. The histogram of the simulated p-values is given in Figure 1 below.
The p-values in \autoref{fig:mar} are clearly not uniformly distributed. Therefore, the Gamma test of \textcite{Marsaglia2004} is not exact and fails to reject far too often, which results in a lack of power or false confidence in a chosen model.

If the test statistic can be standardised in some way so that it is parameter invariant (its value and distribution do not depend on the parameters of the model) then it is usually possible to simulate accurately the distribution of the test statistic under the null hypothesis. This strategy works for location-scale and log-location-scale distributions \parencite[102]{DAgostino1986}. For the Gamma distribution, as a convenient counterexample, it is not possible to eliminate the shape parameter and thus a different approach is needed.

\textcite{Gelman1996} describe a general, Bayesian approach for calculating a posterior predictive p-value (ppp) based on the ideas of \textcite{Rubin1984}. More recently, authors have considered various problems using the methodology of \textcite{Gelman1996}, including multivariate data \parencite{Crespi2009}, discrete data \parencite{Gelman2000}, hierarchical models \parencite{Sinharay2003}, pharmacokinetic models \parencite{Yano2001}, \textit{etc}.

\begin{figure}
\centering
\includegraphics[width=0.8\figwidth]{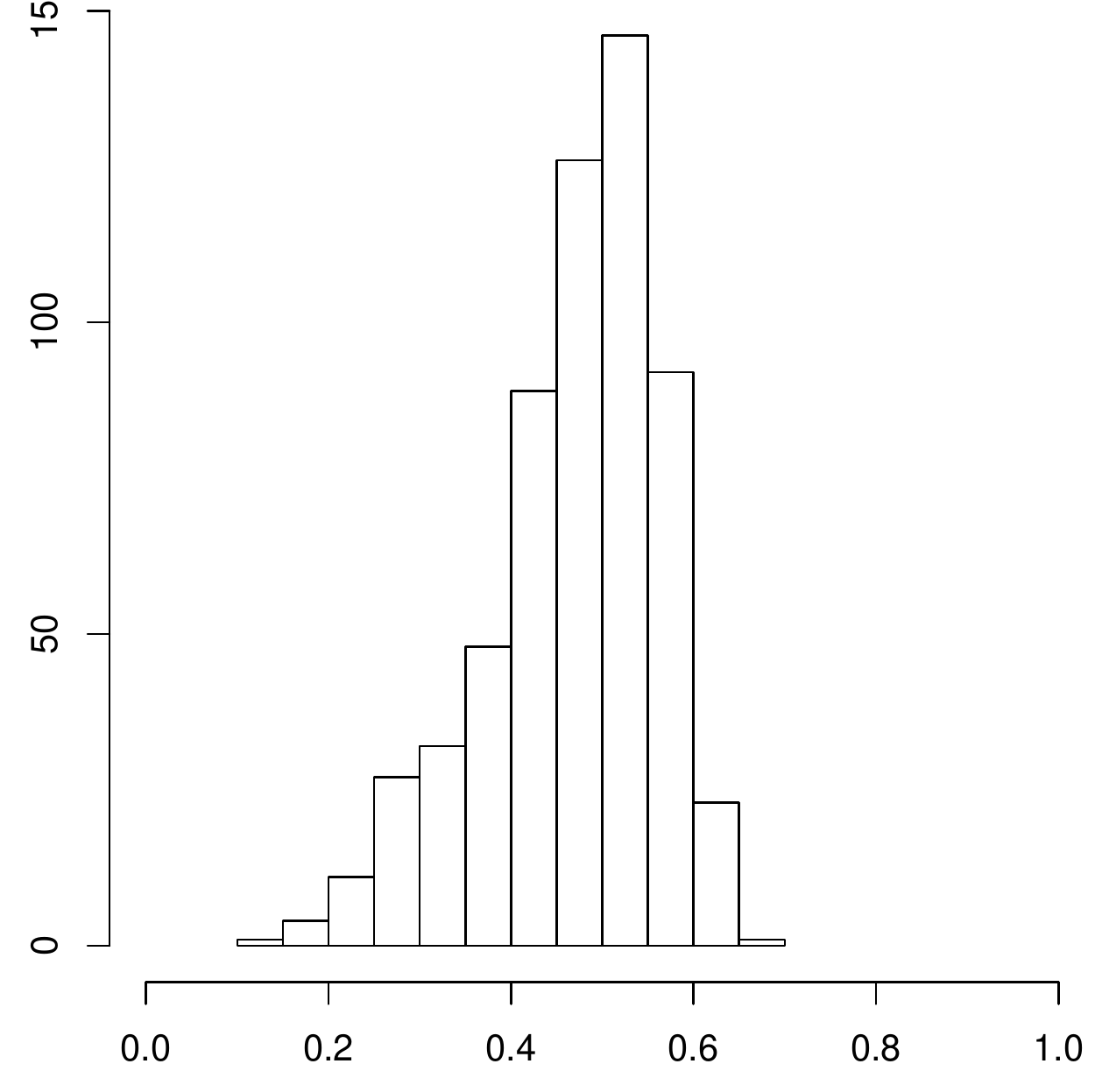}
\caption{Calculated posterior predictive p-values from Gamma samples.} \label{fig:ppp}
\end{figure}

However, the approach of \textcite{Gelman1996} has been criticised --- see, for example, the comment by Rubin on \textcite{Gelman1996}, or \textcite{Bayarri2000}. Of note, when the test statistic chosen depends on the parameters of the model then the resulting test is not exact in the classical sense explained above. To illustrate this we simulated 600 $Gamma(4,8)$ samples of size 12 and calculated the ppp based on the AD statistic for each one. The resulting histogram is given in \autoref{fig:ppp}. It is clear that the p-values are pulled toward 0.5 and require calibration.

This problem of non-Uniform p-values is explained in detail in \textcite{Robins2000}, along with some methods of addressing it asymptotically. Among these methods is what \textcite{Robins2000} refer to as the ``double parametric bootstrap'', which in turn is based on an idea of \textcite{Beran1988}, who called it pre-pivoting. The approach described in \autoref{sec:newmethod} of this paper is a fully Bayesian adaptation of these ideas.

\subsection{Objectives and outline of the present paper} \label{sec:outline}
In this paper the generic problem of testing whether a sample originates from a hypothesized model where some or all parameters of the model are unknown is addressed. A new test is introduced, based on the posterior and posterior predictive distributions, that produces valid p-values in the classical sense. The new test is compared to the parametric bootstrap in two examples:

\begin{enumerate}
\item The first is the independent and identically distributed Gamma observations model (\autoref{sec:gammaimplement}) which is an example where the parametric bootstrap approach works well and we show that the new test performs equally well, both in terms of achieving the target significance level and in terms of power.
\item The second is the independent and identically distributed Generalised Pareto observations model given a known threshold (\autoref{sec:gpdimplement}) where we note that the new test procedure comes much closer to achieving the desired significance level than the parametric bootstrap approach, and as a result, achieves higher power for the same test statistic.
\end{enumerate}

In \autoref{sec:newmethod} we motivate the new test, and present an algorithm for its implementation. The core of the algorithm rests on the idea that in order to arrive at an accurate test we must make full use of all information that can be obtained from the sample. This goal can be achieved through an objective Bayes framework.

In \autoref{sec:discuss} and \autoref{sec:con} we briefly summarise the results of the experiments and give concluding remarks.

\section{New suggested methodology} \label{sec:newmethod}

\subsection{Mathematical motivation} \label{sec:generalmaths}

Let $\mathbf{X}$ be a random variable of dimension $n$ from an unknown distribution, $\mathbf{x}$ an observation of $\mathbf{X}$, (that is, $\mathbf{x}$ is the observed sample), and $m$ a hypothesized model with unknown parameter values $\boldsymbol\theta$. Throughout, we will use bold font to denote vectors. We denote observed quantities with lower case letters and random variables with upper case letters, except for $\boldsymbol\theta$ which we consider to be a scalar or vector random variable throughout. Assume that, after having observed $\mathbf{x}$, the parameter uncertainty is captured in a posterior distribution $p(\boldsymbol\theta|\mathbf{x},m)$ for $\boldsymbol\theta$.
First choose a summary statistic $S$ that compares the sample $\mathbf{X}$ to the model m in a meaningful way; we write $S=S(\mathbf{X}|m)$. In principle, no restriction is placed on the form of the test statistic other than the notion that it should be a function of the sample, and optionally of the parameter values, given a model, and that it should increase as the discrepancy between the sample and the model increases.

In general, in order to calculate $S$ the parameters $\boldsymbol\theta$ need to be specified, as is the case for the KS and AD statistics. Thus, we can usually only calculate the statistic in the form $S=S(\mathbf{X}|\boldsymbol\theta,m)$.

In order to remove the dependence of $S(\mathbf{X}|\boldsymbol\theta,m)$ on $\boldsymbol\theta$ we replace it by its expectation under the posterior distribution $p(\boldsymbol\theta|\mathbf{X},m)$ for $\boldsymbol\theta$ given $\mathbf{X}$. That is, we define $S(\mathbf{X}|m)$ as 
\begin{equation}
S(\mathbf{X}|m)=E_{\boldsymbol\theta}[S(\mathbf{X}|\boldsymbol\theta,m)]=\int S(\mathbf{X}|\boldsymbol\theta,m) p(\boldsymbol\theta|\mathbf{X},m) d\boldsymbol\theta  \label{eq:defS} 
\end{equation}

The statistic $S(\mathbf{X}|m)$ is a random variable, and we can determine its distribution if we can determine the distribution of $\mathbf{X}$. Under a Bayesian approach, given an observed sample $\mathbf{x}$ from model $m$, we work with the posterior predictive distribution of $\mathbf{X}$, namely
\begin{equation} \label{eq:defpredpost}
p(\mathbf{X}|\mathbf{x},m)=\int f(\mathbf{X}|\boldsymbol\theta,m) p(\boldsymbol\theta|\mathbf{x},m) d\boldsymbol\theta
\end{equation}
where $f(\mathbf{X}|\boldsymbol\theta,m)$ is the likelihood implied by the model. In many cases $f$ can be expressed explicitly, but this is not a requirement – being able to simulate from the model as well as from the posterior distribution $p(\boldsymbol\theta|\mathbf{X},m)$ is sufficient for implementing this step.

Given an observed value
\begin{equation} \label{eq:defs}
s(\mathbf{x}|m)=E_{\boldsymbol\theta}[s(\mathbf{x}|\boldsymbol\theta,m)]=\int s(\mathbf{x}|\boldsymbol\theta,m) p(\boldsymbol\theta|\mathbf{x},m) d\boldsymbol\theta
\end{equation}
of the test statistic $S(\mathbf{X}|m)$, we now calculate the probability that $S(\mathbf{X}|m)\geq s(\mathbf{x}|m)$, namely
\begin{equation} \label{eq:defP}
P=P(S(\mathbf{X}|m)\geq s(\mathbf{x}|m))=\int_{S(\mathbf{X}|m)\geq s(\mathbf{x}|m)} {p(\mathbf{X}|\mathbf{x},m) } d\mathbf{X}
\end{equation}

The expression in \autoref{eq:defP} is a p-value that behaves as we would expect from a classic hypothesis test.

The key difference between \autoref{eq:defP} and the corresponding expression in \textcite[Equation 5 on p. 738]{Gelman1996} is the order of integration. In \autoref{eq:defP} every term, namely $p(\mathbf{X}|\mathbf{x},m)$ as in \autoref{eq:defpredpost}, $S(\mathbf{X}|\boldsymbol\theta,m)$ as in \autoref{eq:defS} and $s(\mathbf{x}|\boldsymbol\theta,m)$ as in \autoref{eq:defs}, is first integrated over $\boldsymbol\theta$, using the appropriate posterior distribution. Then, importantly, the desired p-value is obtained by integrating over the predictive posterior distribution for $\mathbf{X}$. \textcite{Gelman1996} integrate a conditional p-value, namely conditional on $\boldsymbol\theta$, over the posterior for $\boldsymbol\theta$.

Vital to understanding this difference is understanding that the posterior distribution $p(\boldsymbol\theta|\mathbf{x},m)$ in \autoref{eq:defs} is not the same as the posterior distribution $p(\boldsymbol\theta|\mathbf{X},m)$ in \autoref{eq:defS}, which is determined in practice based on replicate samples drawn from \autoref{eq:defpredpost}.

Note that we have placed no restrictions on the model so far, other than being able to simulate from the model itself, given parameter values, and being able to simulate from the posterior distribution of the model parameters.

In \autoref{sec:gammaimplement} we will consider a specific model and some of the technicalities that may arise. For example, in many cases the expectation $s(\mathbf{x}|m)=E_{\boldsymbol\theta}[s(\mathbf{x}|\boldsymbol\theta,m)]$ cannot be derived explicitly, and its empirical calculation may be slow. However, the statistic $s(\mathbf{x}|m)$ can be approximated by:
$s(\mathbf{x}|m)=E_{\boldsymbol\theta} [s(\mathbf{x}|\boldsymbol\theta,m)]\approx s(\mathbf{x}|E_{\boldsymbol\theta} [\boldsymbol\theta],m)\approx s(\mathbf{x}|\hat{\boldsymbol\theta}(\mathbf{x}),m)$,
where $\hat{\boldsymbol\theta}(\mathbf{x})$ is an estimate of $\boldsymbol\theta$ based on the sample.
Similarly, $S(\mathbf{X}|m)$ can be approximated as
\begin{equation} \label{eq:approxS}
S(\mathbf{X}|m)=E_{\boldsymbol\theta}[S(\mathbf{X}|\boldsymbol\theta,m)]\approx S(\mathbf{X}|\hat{\boldsymbol\theta}(\mathbf{X}),m)
\end{equation}
When the p-value in \autoref{eq:defP} is calculated through simulation, approximation \ref{eq:approxS} can eliminate the need to obtain the posterior distribution $(p(\boldsymbol\theta|\mathbf{X}^*,m))$ for each draw $\mathbf{X}^*$ of $\mathbf{X}$, which increases execution speed. Both the efficiency and effectiveness of this approximation can differ dramatically from one model to another. It turns out, however, that in the case of the i.i.d. Gamma model the approximation is particularly useful.

\subsection{The parametric bootstrap and posterior predictive check methods} \label{sec:parbootstrap}

The parametric bootstrap is implemented as follows:
\begin{enumerate}
\item Obtain base parameter estimates $\hat{\boldsymbol\theta}$. \label{step:1}
\item Calculate the base statistic $S(\mathbf{x}|\hat{\boldsymbol\theta}(\mathbf{x}),m)$. \label{step:2}
\item Draw $N$ new samples $\mathbf{x}_i,\ i=1,\ldots,N$ from the model $f(\mathbf{X}|\hat{\boldsymbol\theta},m)$ given the parameter estimates from Step \ref{step:1}. \label{step:3}
\item Calculate $N$ new statistics $S(\mathbf{x}_i|\hat{\boldsymbol\theta}(\mathbf{x}_i),m)$ corresponding to each new sample drawn in Step \ref{step:3}. The parameter estimation procedure must be repeated for each new sample. \label{step:4}
\item Calculate the proportion of the new statistics (from Step \ref{step:4}) that exceed the base statistic (from Step \ref{step:2}) and report this result as a p-value. \label{step:5}
\end{enumerate}

The posterior predictive check adapts the parametric bootstrap as follows:
\begin{enumerate}
\item Replace Step \ref{step:1} above with: Simulate $N^*$ sets of parameter values $(\boldsymbol\theta_1,\ldots,\boldsymbol\theta_{N^*})$ from the posterior distribution given the base sample $(p(\boldsymbol\theta|\mathbf{x},m))$.
\item Repeat Steps \ref{step:2} to \ref{step:5} above for each draw $\boldsymbol\theta_i$ to obtain $N^*$ p-values.
\item Average these $N^*$ p-values and report this result as a p-value.
\end{enumerate}

\subsection{Sketch of the new algorithm} \label{sec:simalgorithm}

The proposed new test can be implemented through the following simulation algorithm:
\begin{enumerate}
\item Derive an objective prior for model $m$. \label{step:a}
\item Given the objective prior derived in Step \ref{step:a}, and given an observed sample $\mathbf{x}$, simulate replicate parameters $\boldsymbol\theta_1^*,\ldots,\boldsymbol\theta_N^*$ from the posterior distribution $p(\boldsymbol\theta|\mathbf{x},m)$ for $\boldsymbol\theta$. \label{step:b}
\item Calculate the observed test statistic $s(\mathbf{x}|m)$. If $s(\mathbf{x}|m)$ depends on the parameters $\boldsymbol\theta$ then, using the replicates $\boldsymbol\theta_1^*,\ldots,\boldsymbol\theta_N^*$ simulated in Step \ref{step:b}, calculate $s(\mathbf{x}|m)$ as the average of the statistics $s(\mathbf{x}|\boldsymbol\theta_i^*,m),\ i=1,\ldots,N$ (refer to \autoref{eq:defs}). \label{step:c}
\item Simulate replicate samples $\mathbf{X}^*,\ldots,\mathbf{X}^*$ from the posterior predictive distribution $p(\mathbf{X}|\mathbf{x},m)$ of $\mathbf{X}$. That is, for each replicate parameter $\boldsymbol\theta_i^*,\ i=1,\ldots,N$ from Step \ref{step:b}, simulate a replicate sample $X_i^*$ from the distribution $f(\mathbf{X}|\boldsymbol\theta_i^*,m)$ (refer to \autoref{eq:defpredpost}). \label{step:d}
\item For each replicate sample $X_i^*, i=1,\ldots,N$, calculate the test statistic $S_i^*= S(X_i^*|m)$. If $S(X_i^*|m)$ depends on the parameters $\boldsymbol\theta$ then do Steps \ref{step:i} and \ref{step:ii} below: \label{step:e}
\begin{enumerate}
	\item Given the objective prior derived in Step \ref{step:a}, and given the replicate sample $X_i^*$ from Step \ref{step:d}, simulate parameters $\boldsymbol\theta_{i1}^{**},\ldots,\boldsymbol\theta_{iN}^{**}$ from the posterior distribution $p(\boldsymbol\theta|X_i^*,m)$ for $\boldsymbol\theta$. \label{step:i}
	\item Calculate $S(X_i^*|m)$ as the average of the statistics $S(X_i^*|\boldsymbol\theta_{ij}^{**},m),\ j=1,\ldots,N$ (refer to \autoref{eq:defS}). \label{step:ii}
\end{enumerate}
\item To calculate the p-value, compare the observed test statistic $s(\mathbf{x}|m)$, from Step \ref{step:c}, with its simulated distribution $[S_1^*,\ldots,S_N^*]$ from Step \ref{step:e} (refer to \autoref{eq:defP}). Explicitly, use a continuity adjustment and calculate the p-value as $P=[count(S_1^*,\ldots,S_N^*>s)+0.5]/(N+1)$.
\end{enumerate}

\section{Implementation for the Gamma distribution} \label{sec:gammaimplement}

Consider the following form of the pdf of the Gamma distribution:
\begin{equation} \label{eq:gammadist}
f(x|\alpha,\lambda)=\frac{\lambda^\alpha}{\Gamma(\alpha)}x^{\alpha-1} e^{-\lambda x},\ x>0
\end{equation}

In terms of the notation of the previous section, the parameter vector is $\boldsymbol\theta=(\alpha,\lambda)$.

\subsection{Objective prior distribution} \label{sec:gammaprior}
The maximal data information (MDI) prior \parencite[112--116]{zellner1997} is used as an objective prior. To quote Zellner, the MDI prior provides ``maximal prior average data information relative to the information in the prior distribution''.

Alternative priors such as the Jeffreys prior \parencite{jeffreys1998,Priors} can be used but one must be careful of additional restrictions placed on the parameters. The test procedure may malfunction or fail when the parameters estimated from the sample fall inside or near the restricted area of their domain. The MDI prior does not create such restrictions, which further motivates its use.

The MDI prior is defined as $\exp\{E[\log f(x)]\}$, which works out to:
\begin{equation} \label{eq:MDIpriorGamma}
\pi(\alpha,\lambda)=\frac{\lambda}{\Gamma(\alpha)} e^{(\alpha-1)\psi(\alpha)-\alpha}
\end{equation}
where $\psi()$ is the digamma function.

\subsection{Posterior distribution} \label{sec:gammapost}
Given the MDI prior (\autoref{eq:MDIpriorGamma}) and observations $\mathbf{x}$, the posterior distribution is
\begin{equation} \label{eq:postGamma}
p(\alpha,\lambda|\mathbf{x})\propto \lambda^{(n\alpha+2)-1} e^{-\lambda\sum x_i} \Gamma(\alpha)^{-(n+1) } e^{\alpha\left(\sum \log{x_i} +\psi(\alpha)-1\right)-\psi(\alpha)}
\end{equation}

Therefore, $\lambda|\alpha,\mathbf{x} \sim Gamma(n\alpha+2,\sum x_i )$, and thus
\begin{equation} \label{eq:postGammalambda}
p(\lambda|\alpha,\mathbf{x})=\frac{\left(\sum x_i\right)^{n\alpha+2}}{\Gamma(n\alpha+2)} \lambda^{(n\alpha+2)-1} e^{-\lambda\sum x_i }
\end{equation}
so that
\begin{equation} \label{eq:postGammaalpha}
p(\alpha|\mathbf{x})\propto \Gamma(\alpha)^{-(n+1)} e^{\alpha\left(\sum \log{x_i} +\psi(\alpha)-1\right)-\psi(\alpha)} \Gamma(n\alpha+2) \left(\sum x_i \right)^{-(n\alpha+2)}
\end{equation}

The fastest way to simulate accurately from the posterior distribution (\autoref{eq:postGamma}) appears to be as follows: First simulate values of $\alpha$ from $p(\alpha|\mathbf{x})$ in \autoref{eq:postGammaalpha} and then, given the $\alpha$ values, simulate corresponding values for 
$\lambda$ from $p(\lambda|\alpha,\mathbf{x})$ in \autoref{eq:postGammalambda}. 

\subsection{Test statistic} \label{sec:gammateststats}

Since the object of our comparison is to compare methods of obtaining the distribution of the test statistic and not to investigate or compare the power of statistics, we will focus only on one statistic going forward. We will use the AD statistic as it is well known and has good power. When testing for Normality, which has been heavily studied, the AD statistic (along with the Shapiro-Wilk statistic) has been shown to have high power against the general alternative \parencite{Razali2011}. While less comparisons have been done in the case of the Gamma distribution, we refer to \textcite{Henze2012} who show that the AD statistic has the highest power among the well-known statistics in the case of the Gamma distribution.

Given an i.i.d. sample $\mathbf{x}=(x_1,\ldots,x_n)$ with CDF $F_X (\boldsymbol\theta)$, and corresponding order statistics $x_{(1)},\ldots,x_{(n)}$, the AD statistic $A^2$ is defined as:
\begin{equation} \label{eq:ADstatgeneral}
A^2=s(\mathbf{x}|\boldsymbol\theta)=-n-\sum_{k=1}^n{\frac{2k-1}{n} [\log{F_X(x_{(k)}|\boldsymbol\theta)}+\log (1-F_X(x_{(n+1-k)}|\boldsymbol\theta)) ]}
\end{equation}

Given a specific time constraint, it is possible to achieve higher power for the proposed test using approximation \ref{eq:approxS}, through increased sampling from the predictive posterior distribution. Furthermore, for the sake of speed, parameter estimation is performed throughout using the method of moments.

\subsection{Type 1 error and power of proposed tests} \label{sec:gammatestserrorandpower}

We calculate, through simulation, the type 1 error and power of the proposed tests over a large number of samples from various distributions.

\subsubsection{Type 1 error} \label{sec:gammateststype1error}
First we consider data from Gamma distributions (null hypothesis is true) and determine to what extent each test works as expected from a classical hypothesis test. Specifically, we expect that if a significance level is chosen as $\alpha^*$ (say) then the test will falsely reject the null hypothesis proportion $\alpha^*$ of the time. As stated above, this is equivalent to the p-value being uniformly distributed.

\subsubsection{Power} \label{sec:gammatestspower}
Second we consider data from alternative distributions. We show that the test will correctly reject the null hypothesis more often than proportion $\alpha^*$ (significance level) in all cases. We discuss the effect that failing to achieve the correct significance level as stated in \autoref{sec:gammateststype1error} on power comparisons.

\subsubsection{Design of simulation study} \label{sec:gammatestsdesign}
Only two small sample sizes are used for illustration: 12 and 24. This is to highlight the fact that our proposed approach (\autoref{sec:newmethod}) is non-asymptotic. For each combination of sample size and distribution, 112,000 samples are simulated and the test is performed on each sample independently. The p-values are recorded and summarised in the form of rejection rates.

\begin{figure}
\centering
\includegraphics[width=1.2\figwidth]{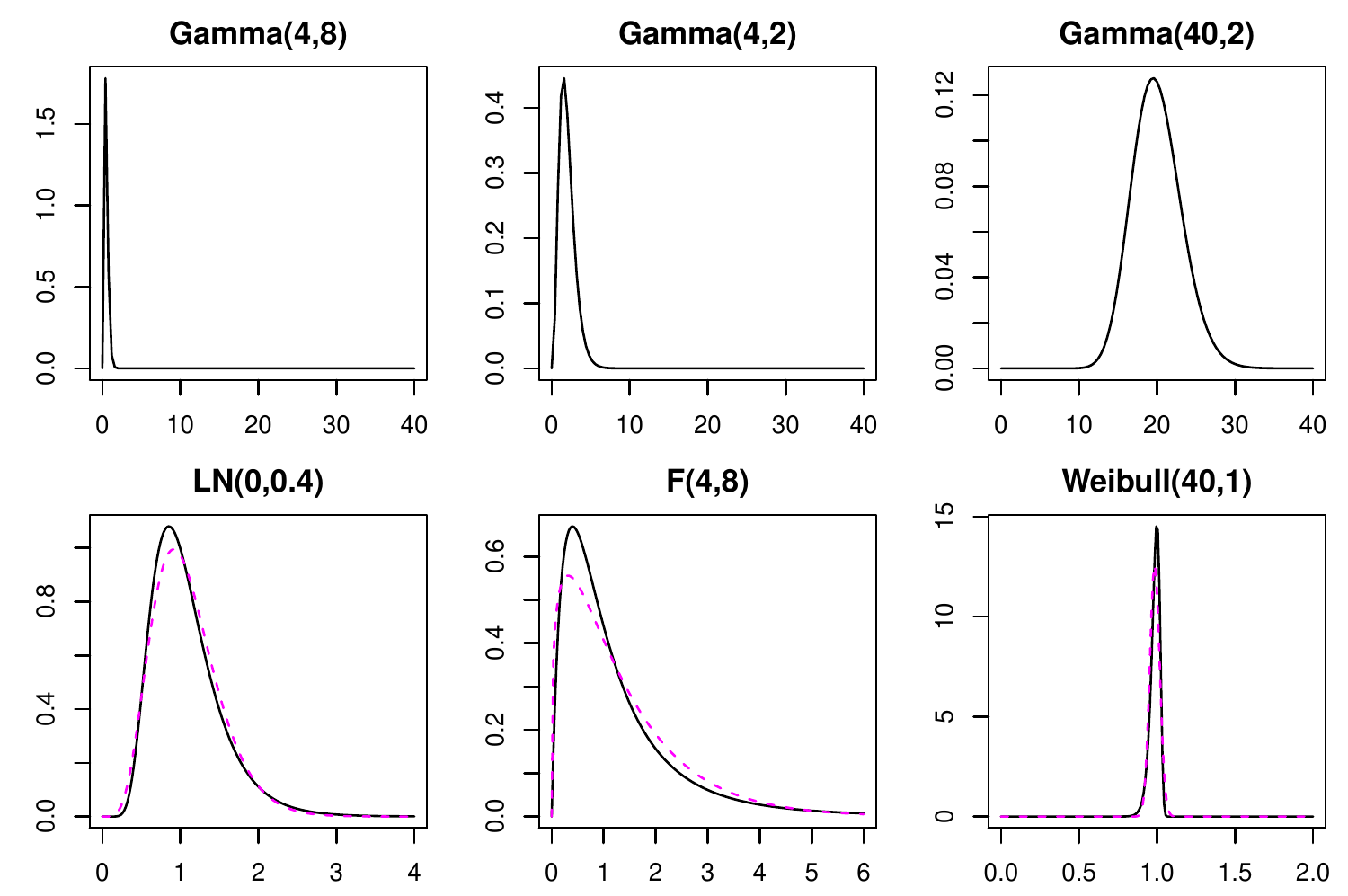}
\caption{Illustration of various distributions used (solid lines) along with Gamma fit (dotted lines).} \label{fig:gammadists}
\end{figure}

The distributions used under the null hypothesis are Gamma(4, 8), Gamma(4, 2) and Gamma(40, 2). The distributions used under the alternative hypothesis are the Log-Normal(0, 0.4), F(4, 8) and Weibull(40, 1). All these distributions are illustrated in \autoref{fig:gammadists} along with the fitted Gamma approximations. The Gamma fit is to highlight the extent to which the alternative distribution differs from the Gamma distribution; note that the Log-Normal distribution is quite close to the best-fitting Gamma distribution.

\subsubsection{Results} \label{sec:gammatestsresults}
The results of the simulation study are summarized in \autoref{tbl:gammarejrates}. Clearly there are no significant differences between the parametric bootstrap method (\autoref{sec:parbootstrap}) and the Bayes method (\autoref{sec:simalgorithm}) as the minor discrepancies at the fourth decimal are all less than one standard deviation under the null hypothesis of Uniform p-values.

\begin{table}
\centering
\begin{tabular}{|l| c| c| }\hline
 Distribution and Method & Sample Size 12 & Sample Size 24 \\ \hline
 G(4,8) Bayes & 0.0493 & 0.0492 \\ \hline
 G(4,8) ParBoot & 0.0496 & 0.0495 \\ \hline
 G(4,2) Bayes & 0.0502 & 0.0513 \\ \hline
 G(4,2) ParBoot & 0.0500 & 0.0513 \\ \hline
 G(40,2) Bayes & 0.0498 & 0.0494 \\ \hline
 G(40,2) ParBoot & 0.0497 & 0.0492 \\ \hline
 LN(0,0.4) Bayes & 0.0666 & 0.0862 \\ \hline
 LN(0,0.4) ParBoot & 0.0666 & 0.0865 \\ \hline
 F(4,8) Bayes & 0.0971 & 0.1572 \\ \hline
 F(4,8) ParBoot & 0.0976 & 0.1576 \\ \hline
 W(40,1) Bayes & 0.1560 & 0.3028 \\ \hline
 W(40,1) ParBoot & 0.1562 & 0.3026 \\ \hline
  \end{tabular}
  \caption{Rejection rates at $\alpha=5\%$ for Gamma tests for different sampling distributions and sample sizes} \label{tbl:gammarejrates}
\end{table}

The reason for the lack of discrepancy is most likely because of the low posterior variance (or accuracy of the parameter estimation). It is for this reason that we now go on to consider a case where the posterior variance is much larger, namely the GPD.

\section{Implementation for the Generalised Pareto Distribution} \label{sec:gpdimplement}

The Generalised Pareto Distribution is used to model the tail (extreme values) of a distribution beyond a given threshold. If the threshold is known we can subtract it from all the observations and consider it to be zero. This makes the GPD a 2-parameter distribution.

\begin{equation} \label{eq:GPDdef}
f(x|\gamma,\sigma)=\frac{1}{\sigma}\left[1 + \frac{\gamma x}{\sigma}\right]^{-\frac{1}{\gamma}-1} , x<-\frac{\sigma}{\gamma}
\end{equation}

We investigate the differences between the parametric bootstrap approach and the Bayes approach with respect to testing the hypothesis that a sample consists of independent GPD observations above a known threshold. The implementation proceeds in the same order as for the Gamma distribution, with only minor differences highlighted in \autoref{sec:MLissues}.

\subsection{Posterior distribution} \label{sec:gpdpost}

We will simulate from this posterior using the Metropolis-Hastings algorithm with proposal $\gamma_c\sim N(\gamma_j,0.05^2)$ and $\log\sigma_c\sim N(\log\sigma_j,0.1^2)$. See \textcite[267--301]{Robert2004} for an in-depth general discussion of this algorithm.

\subsection{Problems with Maximum Likelihood estimation} \label{sec:MLissues}

In all cases we calculate the AD statistic for each replicate sample using the Maximum Likelihood (ML) method as implemented in the evir package in R \parencite{evir}. This approach has the drawback that roughly 0.35\% of the time the parameter estimation fails. In these cases we consider the statistic as missing and ignore it for p-value calculations.

As far as each original simulated sample is concerned, where ML fails we consider the p-value missing for the parametric bootstrap approach but calculate it using the posterior mean in the Bayes approach. We compared these results with the results from dropping these cases entirely and noticed no difference; thus, we can safely assume that this problem has no impact on the outcome of the experiment.

\subsection{Design of simulation study} \label{sec:gpdtestsdesign}

In this case only one sample size was used, namely 24. We chose this sample size to illustrate that the new approach is of most value for smaller samples.

\begin{figure}
\centering
\includegraphics[width=0.8\figwidth]{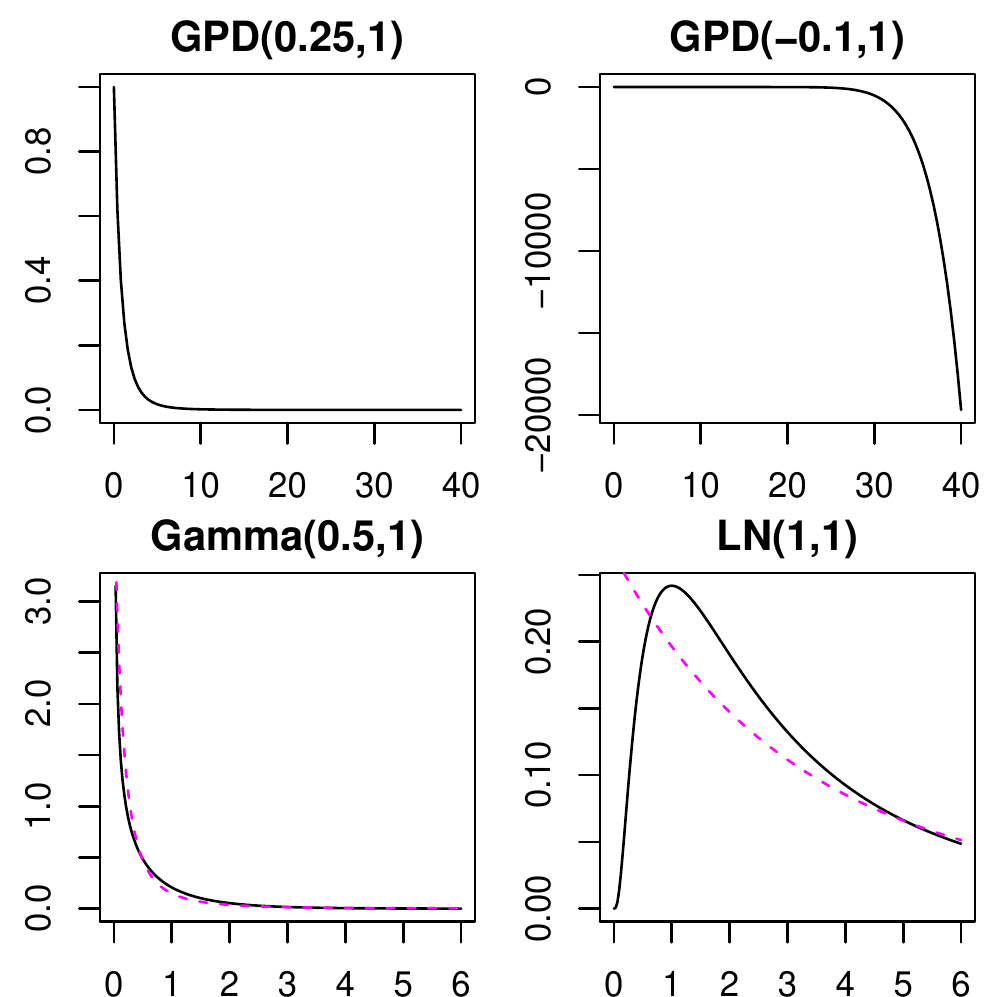}
\caption{Illustration of various distributions used (solid lines) along with GPD fit (dotted lines).} \label{fig:gpddists}
\end{figure}

Null distributions considered are the $GPD(\gamma=0.25,\sigma=1)$ and $GPD(\gamma=-0.10,\sigma=1)$. Alternative distributions considered are the $Gamma(\alpha=0.5,\lambda=1)$ and $Lognormal(\mu=1,\sigma=1)$. These are illustrated in \autoref{fig:gpddists}.

Again we used only the well known AD statistic (\autoref{eq:ADstatgeneral}).

\subsubsection{Results} \label{sec:gpdtestsresults}

\begin{figure}
\centering
\includegraphics[width=\figwidth]{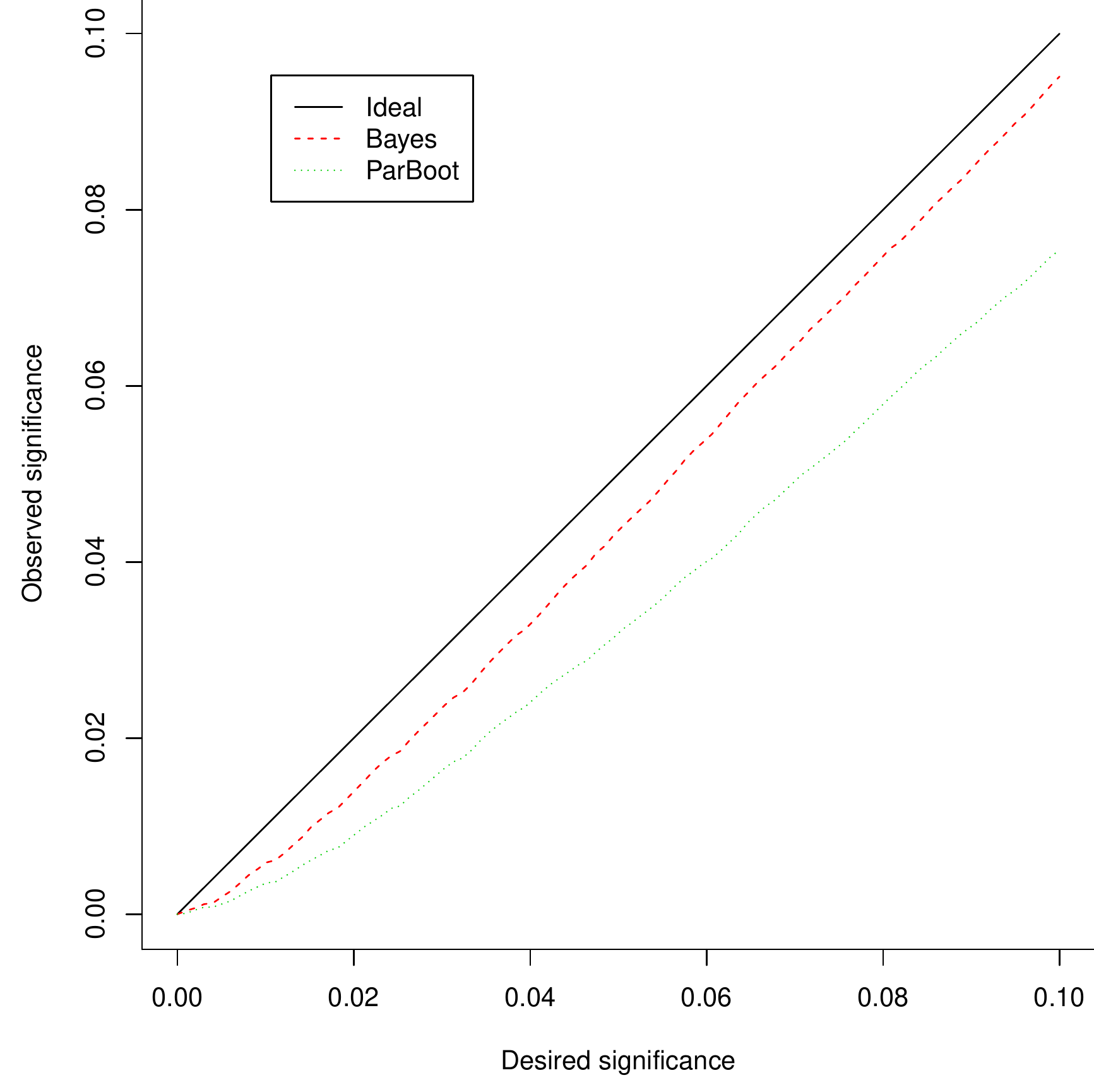}
\caption{Observed coverage of 112,000 GPD tests using AD statistic and sample size 24. Solid line is perfect coverage, dashed line is the Bayes approach and dotted line at the bottom is the parametric bootstrap.} \label{fig:gpdcov1}
\end{figure}

The results of the simulation study are summarized in \autoref{tbl:gpdrejrates}. Here the difference is marked in that the Bayes method (\autoref{sec:simalgorithm}) comes much closer to achieving the desired significance level. This can be seen even more clearly in \autoref{fig:gpdcov1}.

\begin{table}
\centering
\begin{tabular}{|l| c| }\hline
 Distribution and Method & Sample Size 24 \\ \hline
 GPD(0.25,1) Bayes & 0.0432 \\ \hline
 GPD(0.25,1) ParBoot & 0.0316 \\ \hline
 GPD(-0.1,1) Bayes & 0.0385 \\ \hline
 GPD(-0.1,1) ParBoot & 0.0220 \\ \hline
 Gamma(0.5,1) Bayes & 0.4753 \\ \hline
 Gamma(0.5,1) ParBoot & 0.4658 \\ \hline
 LN(1,1) Bayes & 0.1305 \\ \hline
 LN(1,1) ParBoot & 0.1134 \\ \hline
 \end{tabular}
\caption{Rejection rates at $\alpha=5\%$ for GPD tests for different sampling distributions} \label{tbl:gpdrejrates}
\end{table}

\section{Discussion} \label{sec:discuss}

The differences between the observed results of the Gamma experiment and the GPD experiment may be because in the case of the Gamma distribution (especially with large values of the first parameter) the parameter estimation is relatively accurate and straightforward, in stark contrast with the GPD, for which parameter estimation is an open research topic.

The remaining departures of the p-value distribution from the Uniform might be associated with the choice of prior distribution or imperfect posterior simulation. Ideally one would derive a prior distribution such that the test produces perfectly Uniform p-values, but this does not seem mathematically tractable.

\section{Conclusion} \label{sec:con}
It is clear from the results of the simulations that the new test constructed in this paper performs very well and helps address the problem of testing for a distribution with unknown parameters. The new test procedure comes closer to achieving the correct size in a classical hypothesis testing framework, especially for small samples. Furthermore, the power of the test may be higher that what is was in the parametric bootstrap framework and will never be lower. The key difference between the test presented here and previous work is that the Bayesian adaptation works better when faced with problems where the parameter estimation is difficult and carries much uncertainty. The classic approach injects certainty where there is none and this can create false confidence in a chosen model.

\section*{Acknowledgements}
The author wishes to thank Profs Schall, van der Merwe and De Waal as well as Dr van Zyl for asking all the right questions. 

\printbibliography[title=References]

\end{document}